Costantino Sigismondi
*Università di Roma*
*“La Sapienza”*
sigismondi@icra.it

# Impatti lunari: frequenze e monitoraggio

**Abstract:** Lunar impacts have been continuously registered by lunar seismographs from 1969 to 1978, and recently they have been also monitored by a NASA project after several observational campaigns steered by IOTA. Video and naked eye observations, with UTC synchronization, can help to identify impact candidates.

**Introduzione: il progetto NASA sugli impatti lunari**

Coordinato dal Marshall Space Flight Center in Alabama, il progetto *Lunar Impacts*[1] è nato in ambito amatoriale grazie agli sforzi di David W. Dunham e degli osservatori dello IOTA, *International Occultation Timing Association*, sparsi nel mondo [2]. Anche chi scrive ha contribuito a questa ricerca ed al dibattito scientifico con una osservazione ad occhio nudo di un possibile impatto durante l’eclissi totale di Luna del 21 gennaio 2000, osservata da Padova con un binocolo 8x21: l’istante dell’osservazione coincideva con quello osservato da Gary Emerson in America [3]. Questa notizia fu pubblicata in un articolo con Giovanni Paolo Imponente su WGN, il giornale dell’International Meteor Organization e in anteprima su astro-ph [4], dove proponevamo anche un calcolo dell’energia cinetica del meteoroide legata alla luminosità dell’impatto, ed una predizione della frequenza di impatti a partire dall’indice di popolazione (lo “spettro di massa”) delle meteore sporadiche, concetti ripresi in articoli seguenti [8, 9, 10]. Molto da quel tempo è stato fatto, ma ancora ampio è il contributo che gli appassionati possono dare, anche senza strumenti troppo sofisticati. *Get the most scientific value from your telescope and video equipment!* Questo è uno degli spot sul sito dello IOTA dedicato agli impatti lunari [2], che ha dato il via alla campagna della NASA, di cui riporto integralmente lo scopo della missione: “Usare osservazioni a Terra della porzione oscura della Luna per stabilire le frequenze e le dimensioni dei grandi meteoroidi (oltre i 500 grammi) che colpiscono la superficie lunare”. In vista della possibile installazione di basi lunari permanenti alla fine del prossimo decennio “gli astronauti dovranno stare sulla Luna per lunghi periodi di tempo, ed i meteoroidi, con i loro ejecta quando questi creano un cratere da impatto, fanno parte dell’ambiente lunare.”

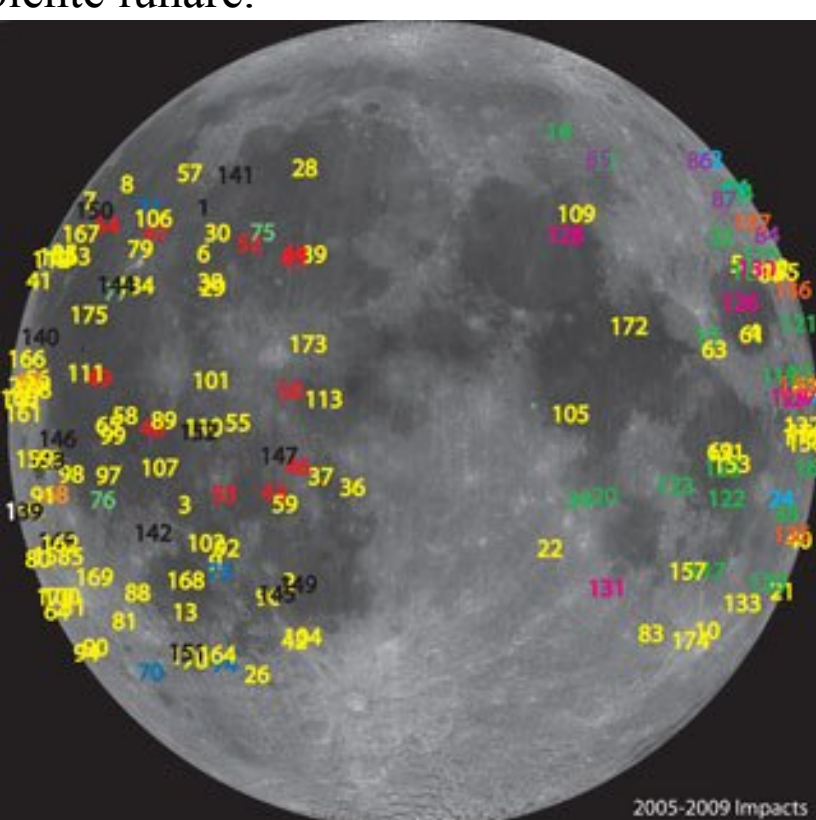

Figura 1. Impatti lunari all'8.7.09: dati NASA-MEO [5].

**Sismografi sulla Luna e meteore sporadiche**

I sismografi installati sulla Luna dagli astronauti delle missioni Apollo hanno registrato tra il 1969 ed il 1978 diversi eventi di impatti [6]. Gli sciami meteorici noti [7], contrariamente a quanto si possa credere, non contribuiscono in modo significativo ai grandi impatti, mentre sono le meteore sporadiche a fornire i corpi, fino a qualche tonnellata, responsabili degli impatti più forti [5,6]. Ad esempio l’impatto del 13.5.1972 vicino alla stazione sismografica di Apollo 14 è stato prodotto da un corpo di 1100 kg ad una velocità di 22.5 km/s [6] mentre uno che ha rilasciato un’energia quattro volte superiore è avvenuto il 17.7.1972 più lontano dalla rete di sismografi. Le masse degli impattori variano tra 100 g ed 1 tonnellata.

L’ipotesi è che una frazione prossima al 100% dell’energia cinetica incidente si trasformi in luce per effetto dell’esplosione del meteoroide sul suolo lunare.

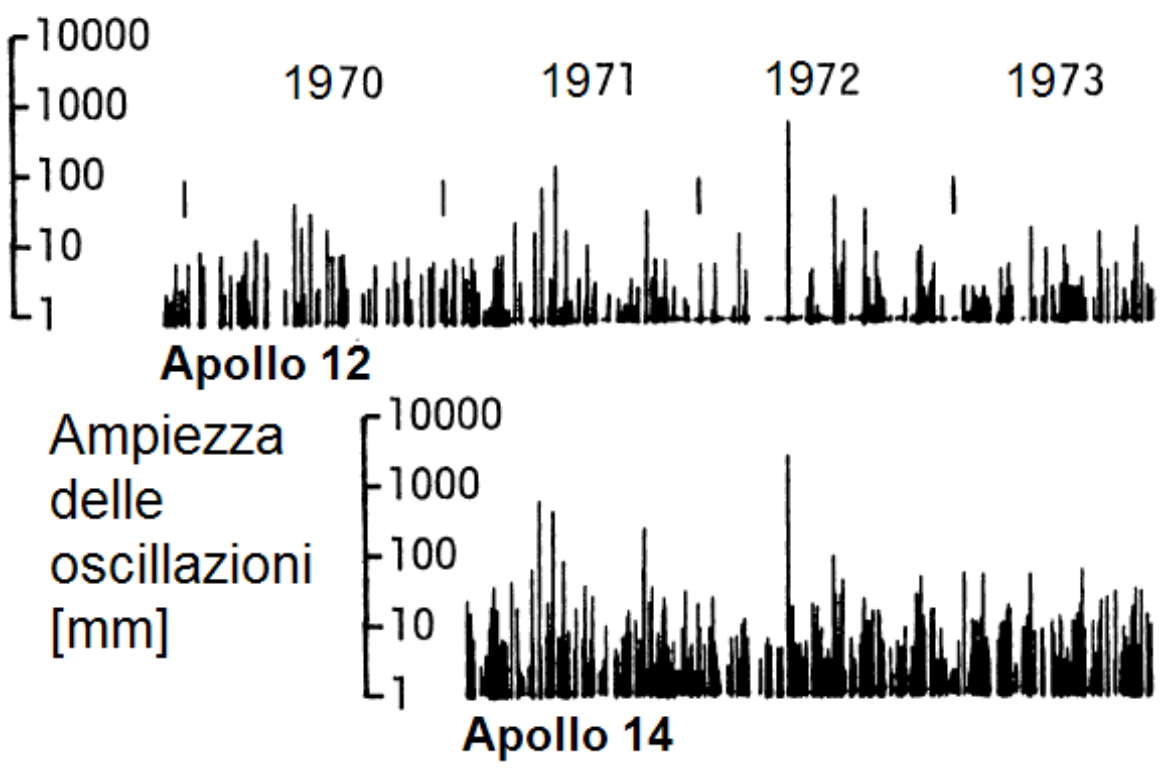

Figura 2. Adattata da [6], si vedono le ampiezze delle oscillazioni dei sismografi lunari. La maggiore supera il metro, registrata nei quattro siti di Apollo 12, 14, 15 e 16.

Da Terra le meteore sporadiche [11] hanno un tasso orario zenitale di circa 10, e sono presenti tutto l’anno. L’osservazione sistematica del cielo per identificarne le più brillanti è possibile e raccomandabile anche nelle notti di Luna piena. Poiché le meteore davvero interessanti hanno magnitudini negative, non esiste nessuna notte sfavorevole alla loro osservazione, tranne quelle nuvolose.

Quanto agli impatti lunari, la parte illuminata della Luna impedisce di registrare eventi che hanno magnitudine superiore alla decima [1,2].

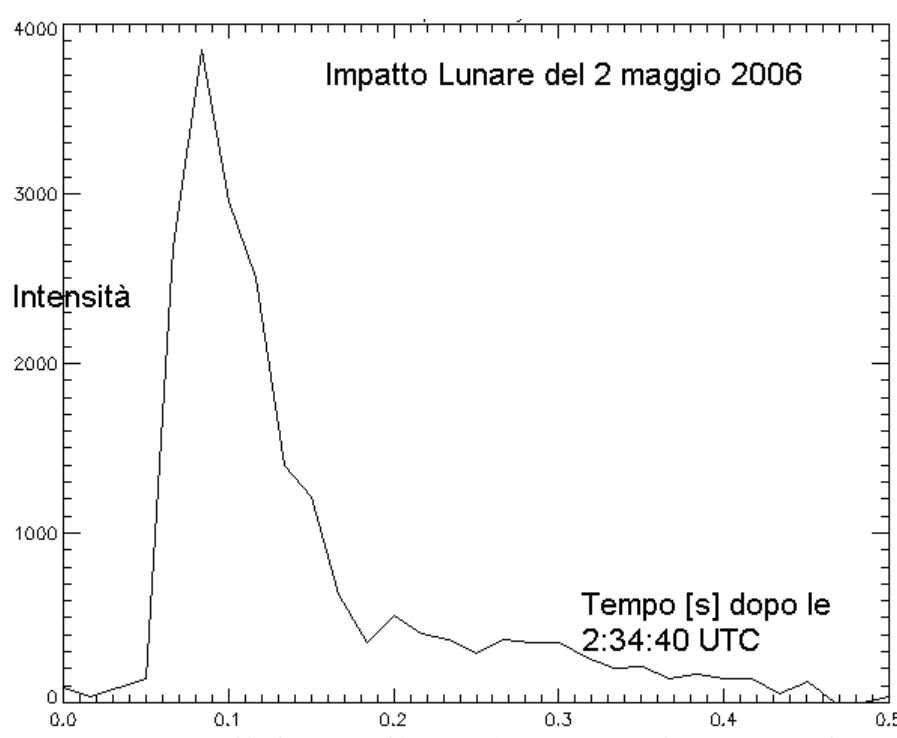

Figura 3. Curva di luce di un impatto lunare, da [12]. Un modello di *swarm* non spiega in modo soddisfacente la durata di questo flash: per effetto mareale il meteoroide iniziale si è disgregato in una piccola nube [*swarm*], ma il tempo di impatto sarebbe comunque molto minore di quello osservato, date le alte velocità [km/s] e le piccole dimensioni [m] in gioco. Anche un modello di raffreddamento sia del suolo stesso che della palla di fuoco, a seguito dell'esplosione del meteoroide avvenuta sul suolo lunare non spiega flash così lunghi. Anche l'angolo di impatto e la natura del suolo possono influire molto sulla durata del flash.

**La formazione del cratere Giordano Bruno**

L'evento descritto nelle cronache [13] di Gervaso di Canterbury (1141-1210), e di seguito in traduzione, potrebbe essere stato quello che ha dato origine al cratere Giordano Bruno [14] di coordinate selenografiche 35.9°N, 102.8°E. La data è la domenica 18 giugno 1178, calendario Giuliano, compatibile con il passaggio di uno *swarm* della cometa di Encke. [15]

"In questo anno, nella domenica prima della festa di S. Giovanni Battista, dopo il tramonto, quando la Luna era appena diventata visibile un fenomeno meraviglioso è stato osservato da cinque o più uomini che stavano seduti guardando la Luna. Ora c'era una Luna nuova brillante, e, come è usuale in questa fase, i suoi corni erano orientati verso Est, ed improvvisamente il corno superiore si è diviso in due. Dal punto centrale della divisione è venuta fuori una torcia fiammeggiante gettando fuori, a distanza considerevole, fuoco, carboni ardenti e lampi. Frattanto il corpo della Luna che era sotto contorta, come in ansia, e, per renderlo nelle parole di coloro che mi hanno riportato ciò per averlo visto con i loro occhi, la Luna pulsante come un serpente ferito. Dopo ritornò al suo proprio stato. Questo fenomeno si ripeté una dozzina di volte o più, la fiamma assumendo vari profili contorcendosi a caso e poi tornando al normale. Quindi dopo queste trasformazioni la Luna da corno a corno, cioè lungo la sua intera lunghezza, prese un aspetto nerastro. Il sottoscritto ricevette questa testimonianza da uomini che videro ciò con i loro occhi, pronti a giocarsi il proprio onore su giuramento che non hanno fatto aggiunte o falsificazioni al loro racconto.
Gervaso di Canterbury"

I calcoli, svolti al sito di calsky.org il 4 giugno 2012 per la domenica 18 giugno 1178 mostrano come effettivamente quel giorno la Luna fosse visibile a 3 gradi dall'orizzonte, per mezz'ora dopo il tramonto. J. Meeus [16] e B. Schaefer [17] confutano questi calcoli, negando la visibilità della Luna e l'attendibilità del cronista. Un'interpretazione alternativa del fenomeno, suggerita dal referee di questo articolo, è la rifrazione irregolare dell'atmosfera terrestre, come si osserva talvolta dalla forma del disco solare presso l'orizzonte. [18]

**Frequenze di impatti in funzione della loro energia**

L'energia cinetica incidente E solo in parte si trasforma in luce. L'efficienza luminosa è proprio la frazione $\eta$ di E che va in luce. Si ritiene che questo valore sia dell'ordine di $\eta=2\cdot10^{-3}$, ma spesso negli articoli occorre invocare valori anche maggiori $\eta\approx10^{-2}$ per rendere ragione delle luminosità osservate.

Nella figura seguente alcuni dati sull'energia del corpo impattante sono calcolati dai parametri cinematici, altri dalle luminosità osservate e, nonostante ciò una legge di potenza descrive bene l'insieme di tutti gli impatti.

L'unità di misura scelta dagli autori del grafico [19] per l'energia cinetica è il kiloton corrispondente a $4.184\cdot10^{12}$J. Un evento come quello di Tunguska del 1908 che è stato di 10 Megaton può accadere in media una volta ogni mille anni [20].

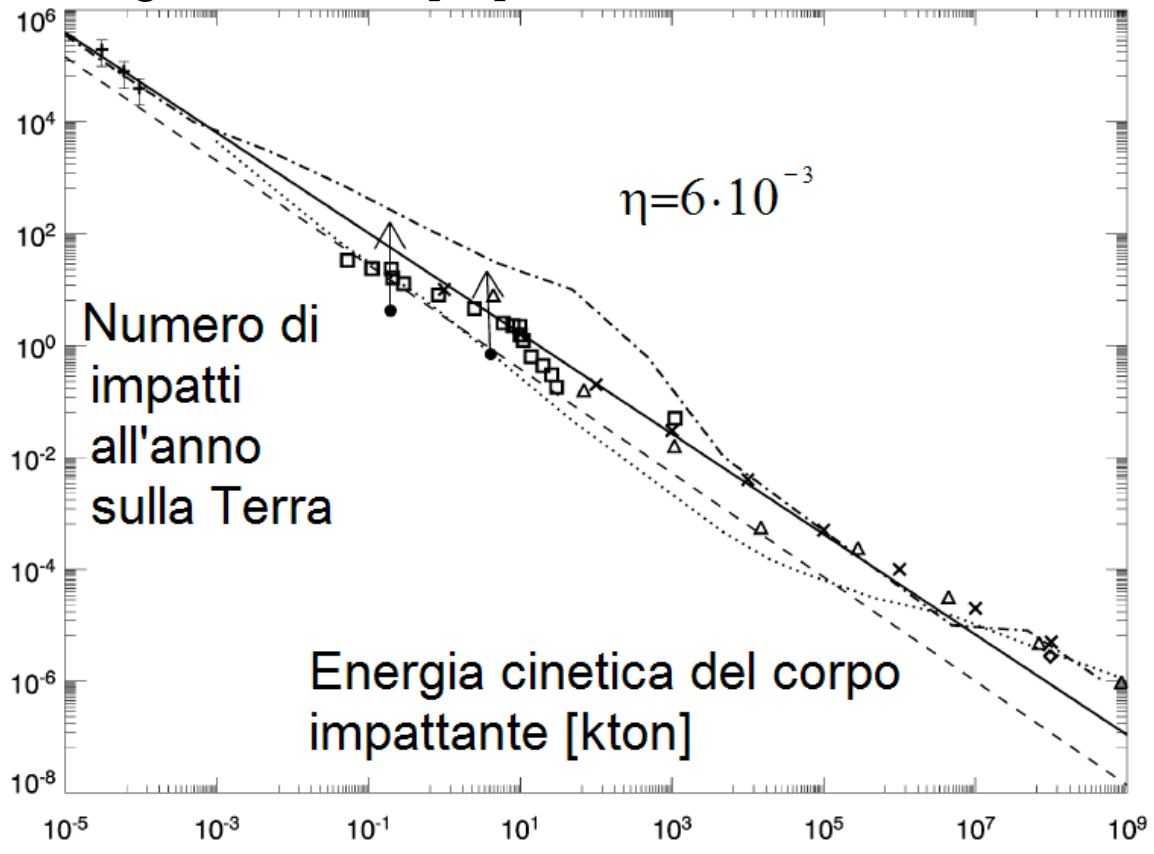

Figura 4. Adattata da Ortiz et al. [19] mostra la legge di potenza che descrive meglio tutti gli impatti lunari e terrestri opportunamente riscalati per diverso coefficiente gravitazionale e le diverse aree in gioco. La Terra ha una maggior forza di gravità della Luna e questo aumenta di 1.3 volte il numero degli impatti per unità di area. Da questo grafico la formula che descrive meglio il numero di impatti N per anno attesi su tutta la Terra per una data energia cinetica E in kton è $N\approx10\cdot E^{-0.9}$ (linea continua), che per Tunguska dà 1 evento ogni 400 anni. La linea tratteggiata è la stima di Brown et al. [20] con pendenza maggiore $N\approx6\cdot E^{-0.95}$, consistente con un evento come Tunguska ogni 1000 anni.

**Conclusioni**

Rileggendo l'eccezionale cronaca di Gervaso di Canterbury, si vede come la Luna rimanga soggetto di continua craterizzazione, anche se i grossi calibri sono molto rari.

La descrizione della Luna rimasta più oscura dopo l'impatto può essere dovuta ad un'atmosfera transiente sulla Luna stessa, soggetto, questo, di studi molto recenti [21].

La frequenza degli eventi ripresi con telescopi da Terra nel progetto della NASA è stata valutata pari a 400 per anno, su tutta la superficie lunare, che è grande quanto l'Africa (38 milioni di km²), la massa degli oggetti è dell'ordine del chilogrammo. Sulla Terra tali oggetti si distruggerebbero in atmosfera dando luogo a bolidi.

L'osservazione di un impatto lunare richiede, per la validazione, la possibilità di fare riprese ben sincronizzate con il tempo universale UTC della parte in ombra della faccia della Luna. È possibile anche osservare pennacchi sul bordo lunare [22]: la loro origine può essere dovuta anch'essa ad impatti di meteoroidi, così come lo è il famoso LTP, *lunar transient phenomenon*, del 15 novembre 1953 2 UT, fotografato da Leon Stuart e che fu valutato di magnitudine -1 e persistette per circa 8 secondi. La sonda Clementine ha trovato un cratere molto recente nella zona, ma a 30 km dal luogo indicato dalla foto. Il caso non è ancora chiuso.[23]

## Bibliografia